\def\galex{{\sl GALEX}}
\def\hst{{\sl HST}}
\def\acs{ACS}
\def\wfc3{WFC3}
\def\spitzer{{\sl Spitzer}}
\def\irac{IRAC}
\def\mips{MIPS}
\def\herschel{{\sl Herschel}}
\def\pacs {PACS}
\def\spire{SPIRE}
\documentclass[12pt,preprint]{aastex}
\usepackage{epsf}
\usepackage{color}
\usepackage{graphicx}
\usepackage{graphics}
\usepackage{epsfig}
\usepackage{longtable}
\usepackage{multirow}
\begin{document}
\title{High-Resolution Mapping of Dust via Extinction in the M31 Bulge}
\author{Hui Dong$^{1,2}$, Zhiyuan Li$^{3,4}$, Q. D. Wang$^{5,3}$, Tod R. 
  Lauer$^2$, Knut A. G. Olsen$^2$, Abhijit Saha$^2$, Julianne J. Dalcanton$^6$,
  Brent A. Groves$^7$}

\affil{$^1$ Instituto de Astrof\'{i}sica de Andaluc\'{i}a (CSIC), Glorieta de la Astronom\'{a} S/N, 18008 Granada, Spain}\affil{$^2$ National Optical Astronomy Observatory,
Tucson, AZ, 85719, USA}\affil{$^3$ School of Astronomy and Space Science, Nanjing University, Nanjing, 210093, China}\affil{$^4$ Key Laboratory of Modern Astronomy and Astrophysics at Nanjing University, Ministry of Education, Nanjing 210093, China}\affil{$^5$ Department of Astronomy, University of Massachusetts,
Amherst, MA, 01003, USA}\affil{$^6$ Astronomy Department, University of Washington,
Seattle, WA, 98195, USA}\affil{$^7$Max Planck Institute for Astronomy,
K$\ddot{o}$nigstuhl 17, D-69117 Heidelberg, Germany}\affil{E-mail: hdong@iaa.es}

\begin{abstract}
We map the dust distribution in the central
180\arcsec\ ($\sim$680 pc) region 
of the M31 bulge, based on \hst\ \wfc3\ and \acs\ observations in ten
bands from near-ultraviolet (2700 \AA ) to
near-infrared (1.5 $\mu$m). This large wavelength coverage gives us great leverage to
detect not only dense dusty clumps, but also diffuse dusty molecular gas. 
We fit a 
pixel-by-pixel spectral energy distributions to 
construct a high-dynamic-range extinction map 
with unparalleled
angular resolution ($\sim$0.5\arcsec , i.e., $\sim$2 pc) and
sensitivity (the extinction uncertainty, $\delta$$A_V$$\sim$0.05). 
In particular, the data allow to directly 
fit the fractions of starlight obscured by individual dusty clumps, and hence 
their radial distances in the bulge. Most of these clumps seem to be
located in a 
thin plane, which 
is tilted with
respect to the M31 disk and appears face-on. 
We convert the extinction map 
into a dust mass surface density map and compare it   
with that derived from the dust emission as observed by 
\herschel . The dust masses in these two maps are consistent with each
other, 
except in the low-extinction regions, where the mass inferred from the
extinction tends to be underestimated. Further, we use simulations to 
show that our method can be used to measure the masses of dusty 
clumps in Virgo cluster early-type galaxies to an accuracy within a factor of $\sim$2. 
\end{abstract}

\section{Introduction}
Dust is an important ingredient for star formation, is often released into the 
interstellar medium (ISM) at the end of stars' lives, absorbs/scatters
light, and radiates significantly from millimeter to
infrared. Therefore, 
accurately mapping the dust 
in the Universe is critical to correcting 
for the foreground extinction, measuring the star 
formation rate, and in turn understanding the evolution of galaxies. 
Radio observations are a widely 
used tool to trace various molecules, from which one may infer the 
dust, assuming a dust-to-gas mass ratio. Because molecular
hydrogen (H$_2$) is notoriously difficult to detect,
carbon monoxide molecule (CO) is typically used to
infer the mass of molecular clouds.
Single-dish radio observations with the high
sensitivity needed to detect diffuse molecular
clouds of relatively low density, however, often suffer from poor angular
resolution at extragalactic distances. 
Interferometry provides high angular resolution, but typically lacks the
sensitivity to detect anything beyond dense, giant molecular cloud
cores. In addition, the relationship between CO intensity and H$_2$ mass
 in different environments is also poorly
understood~\citep[e.g.,][]{bol13,san13}. Alternatively, 
thermal far-infrared (FIR) observations 
can be used to directly detect radiation 
from dust. However, the angular
resolutions of such observatories are poor (even with 
\herschel ,
$\geq$6\arcsec ). Both emissivity and temperature distributions  
 of dust are poorly known in many circumstances. As a result, the inferred ISM 
 mass can be greatly uncertain (Shetty et al. 2009a, 2009b).
 
 Fortunately, dust can be detected  
 through its extinction against 
background starlight. In the Milky
 Way and Magellanic Clouds, the hydrogen 
 column densities of molecular clouds have been 
statistically measured from the 
 near-infrared (NIR) colors of background stars~\citep[][and
 references therein]{lom09,dob08,dob09}. Compared 
 to the radio and far-infrared observations, this method can reach unparalleled 
angular resolution and sensitivity because of the availability of
high-quality NIR source catalogues, such 
as 2MASS~\citep{skr06}. Now the method can be applied to 
galaxies beyond
the Milky Way and Magellanic Clouds, e.g., the M31
disk \citep{dal15}, when starlight can be
resolved into individual 
stars and/or their colors can be utilized 
(e.g.,~\citealt{cro11}). With modern observations in the wavelength
range from IR to
ultraviolet (UV) (e.g., from the Hubble Space Telescope, \hst ), one can, in 
principle, map the interstellar dust at sub-arcsecond resolutions 
and over a broad dynamic range.


The interpretation of extinction mapping depends on the
line-of-sight location of dusty clouds relative to the stellar light 
distribution. Only the starlight emitted in regions behind 
the cloud would suffer from extinction and reddening. Thus, without accounting 
for this effect (i.e., simply assuming that the cloud is in the
foreground), one   
would underestimate the dust column density and mass~\citep{cal94}.  
For example,~\citet{kre13} find that in eight
nearby galaxies, the dust mass derived from the stellar continuum
(without taking into account the partial extinction) is 
 systematically less than that estimated from the FIR emission from \herschel\
 observations. Therefore, determining the obscured fraction of the 
starlight is a key for a successful use of the extinction method.

The work presented here provides a test of the method developed
in~\citet{don14,don15} to address the above problem, using multi-wavelength 
\hst\ data of the M31 bulge. The basic idea 
is that the fraction of the starlight arising from beyond 
an obscuring cloud can be measured together with its extinction 
as a function of wavelength. Therefore, we can use a 
pixel-by-pixel spectral energy 
distribution (SED) fitting to derive an extinction map, and subsequently construct a 
high-resolution dust mass surface density map with a large dynamic range, 
assuming a dust mass to $A_V$ ratio. The inferred obscured fraction of
starlight can be converted to the radial location of the cloud, if the
intrinsic starlight 
distribution and a screen geometry can be reasonably modelled (see \S\ref{s:analysis}). 
Specifically, we utilize ten images 
taken by the \hst\ Wide Field Camera 3 (\wfc3) and
Advanced Camera for Surveys (\acs ) cameras (listed in
Table~\ref{t:obs}) from the near-UV (2700 \AA ) to near-IR (1.5 $\mu$m). The high angular 
resolution (Full Width Half Maximum, FWHM, $\leq$0.15\arcsec , $\sim$0.6
pc) of this set of images, combined with the proximity of M31 enables us to resolve dense dusty
clumps~\citep{ber07}. As a result, the spatial variation of
extinction, due to the clumpy effect of these 
dusty clumps in individual resolution elements, should
be largely reduced. The derived surface density map can also be compared with that produced from the \herschel\
observations to examine the reliability of our method. 


During the past decade, the interstellar 
medium in M31, the closest external massive 
spiral~\citep[$\sim$780 kpc,][]{mcc05},  has been extensively
mapped with various instruments 
at various bands, such as the Westerbork Synthesis Radio Telescope array 
for the \ion{H}{1} 21 cm emission line~\citep{bra09}, IRAM 30 m
telescope for $^{12}$CO~\citep{nie06}, 
\spitzer\ Space Telescope at mid-IR~\citep[3 to 160
$\mu$m,][]{bar06,gor06} and \herschel\ 
Space Telescope at FIR~\citep[100-500 $\mu$m,][]{fri12,smi12} for dust. Much of 
the atom or molecular gas is found in so-called the 
10 kpc star-forming ring. There are moderate 
amounts of  ionized and molecular gas, as well as dust 
in the circumnuclear region (CNR)
 of M31~\citep{bra09,nie06}. For example, based on 
Spitzer/IRAC observations, ~\citet{blo06}
identify a ring-like dusty structure with an extent of 
1 kpc ($\sim$260\arcsec ) in the \spitzer/\irac\
observations, which is part of spiral-like patterns, 
perviously known as the `nuclear spiral'~\citep{jac85}. Through SED fitting of
\spitzer/MIPS data (24-160 $\mu$m),~\citet{li09} report 
$\sim$9$\times$10$^3$ M$_{\odot}$ dust within this nuclear spiral.
~\citet{gro12} give a total dust mass,
$\sim$1.5$\times$10$^5$ M$_{\odot}$, within
the central 1 kpc of the M31 bulge, through analyzing recent \herschel\
observations. With very deep
observations of CO(1-0) and CO(2-1) taken by the 
IRAM-30m telescope,~\citet{mel13} give a lower limit of 4.2$\times10^4$
M$_{\odot}$ molecular gas within the central 30\arcsec\
($\sim$100 pc) of the M31 bulge. 

Both the content and the morphology of the dusty material 
in the CNR of M31 are of great interest to the understanding 
of its origin and interplay with the nuclear environment.
~\citet{blo06}
notice that both the 1 kpc dusty ring and the well-known 10 kpc star forming
ring~\citep[][and references therein]{gor06} are off-center with
respect to M31*, the 
supermassive black hole, a phenomenon reminiscent of that observed in 
the Cartwheel 
galaxy. These rings are suggested to be the result of a head-on collision between
M31 and a satellite galaxy, presumably the nearby M32, 
roughly 200 Myr ago, which 
triggered a density wave propagating through the M31 disk. 
Furthermore, \citet{mel11} find that several dusty clumps in the northwest of the M31
bulge show distinct velocity components, which may represent the 1 kpc
dusty ring and its distortion, due to the putative 
collision. \citet{mel13} also suggest that the extinction 
produced by the 1 kpc dusty ring and its distortion are 
needed to explain the position-angle distribution of planetary nebulae
observed in the central 200\arcsec\ radius of M31. 
However, without knowing the actual spatial locations of the individual
dusty clumps, one cannot be confident about 
their overall structure and the origin.





The rest of the paper is organized as follows. We 
describe the observations and
data reduction in \S~\ref{s:observation}. We present our
pixel-by-pixel SED fitting method, detailed
analysis steps and the results in \S~\ref{s:analysis}. 
We discuss the implications of our results in \S~\ref{s:discussion} and
give a summary of the work in \S~\ref{s:summarize}. 

\section{Data}\label{s:observation}
\subsection{\hst\ multi-band dataset}
Detailed description about the \hst\ multi-band images are
given in~\citet{don14,don15}. Here, we briefly describe the data and
major 
reduction steps, as well as an adaptive smoothing procedure performed on the images. 

Our images were taken in ten bands. Six of them are from the Panchromatic Hubble Andromeda Treasury
program~\citep[PHAT,][]{dal12,wil14}, including F275W (2704 \AA ), F336W (3355 \AA ), F475W (4747 \AA ), 
F814W (8057 \AA ), F110W (1.1 $\mu$m) and F160W (1.5 $\mu$m). 
The other four images are from Program
GO-12174 (Li et al. 2016, in preparation), 
including F390M (3897 \AA ), F547M (5447 \AA ), 
F665N (6656 \AA ) and Program 10006, Program 10760 and
 Program 11833~\citep{wil05} on F435W (4319 \AA ). 
 The F547M band is approximately  
 the traditional Johnson $V$ band, because 
 they have similar central wavelengths. The images in these 
ten bands have different fields-of-view. 

We performed the basic calibration steps, such as identifying 
bad pixels, dark subtraction and flat-fielding correction with
{\tt CALWFC3} and {\tt CALACS} in {\tt PyRAF}\footnote{`{\tt PyRAF}' and `{\tt PyRAF}' are the product of 
the Space Telescope Science Institute, which is operated by AURA for
NASA.}. We used the {\tt Astrodrizzle} to merge
dithered exposures for each pointing and filter. We
corrected for the relative astrometry and
bias offset among pointings of individual filters with the 
method developed by~\citet{don11}. These pointings
were then aligned to the absolute astrometry defined by bright objects found in the 2MASS
catalog~\citep{skr06}. We merged the images
of different pointings into the 250\arcsec$\times250$\arcsec\
(950 pc$\times$950 pc) mosaic for each of the ten bands. 
We converted the image units from the count rate (electron/s) to
intensity (erg cm$^{-2}$ s$^{-1}$ \AA\ $^{-1}$) by multiplying the
mosaics with the `PHOTFLAM' values listed in
Table~\ref{t:obs}. The intensity 
uncertainty of each pixel ($\sigma_n$, where the subscript indicates the
filter number) was derived in~\citet{don15} and 
consists of two parts: statistic and
systematic errors. The statistic error, the intensity variation
among pixels, was empirically derived within the
mosaic, while the systematic uncertainty was from 
`PHOTFLAM' (Table~\ref{t:obs}; see~\citealt{don15} for 
details).  Foreground Galactic extinction~\citep[E(B-V)=0.062, 
i.e., $A_V$=0.17,][]{sch11} was corrected 
for the intensity in each band. We then employed {\tt PSFMATCH} in {\tt PyRAF} to 
construct PSF kernels and convolved them with mosaics one-by-one to 
the resolution of the F160W band. We also rebinned 
the mosaics to the common pixel size (0.13\arcsec) of the WFC3/IR.

We adaptively smoothed the mosaics to make sure that each resolution
element has at least 10$^5$ M$_{\odot}$, in order to minimize the 
systematic errors on age and mass 
due to stochastic fluctuations in the stellar evolutionary synthesis~\citep{fou10}.
 The mass-to-light ratio (M/L) of
the F160W band is the least sensitive to the extinction, 
stellar age and metallicity in our dataset. We thus estimated 
the stellar mass surface density distribution from the intensity in
the band using a conversion M/L$_{F160W}$=1.8, 
where M and L$_{F160W}$ are in units of M$_{\odot}$ and
L$_{\odot}$, appropriate for a 10 Gyr old stellar population with 
solar metallicity. We employed a smoothing square
box\footnote{In principle, one may employ circular smoothing kernels, such as Gaussian or circular top-hat. We have adopted boxy kernel for simplicity. We find no artifacts due to our choice in the final extinction map.} centered on 
each pixel and adaptively scaled to enclose a total of at least 10$^5$ M$_{\odot}$. The average intensity 
and its variance are measured  within this
box for each of the ten \hst\ filters. These measurements are used for the SED-fitting of this 
pixel. The angular resolution of our extinction map is thus 
determined by the size of the box, which generally increases with the 
galactocentric distance. But the box sizes for 93\% of the pixels are still $<$7 pixel,
or 0.9\arcsec . On average, the angular resolution of our final
extinction map is $\sim$ 0.5\arcsec\ or 1.9 pc at the distance of M31.   

The final images in the F336W, F435W, F547M and F160W bands are presented in
Fig.~\ref{f:mosaic}. We can clearly see the
dusty clumps in the CNR as dark fuzzy filaments in the UV band (F336W), which
appear weak in the optical (F435W/F547M) images and finally disappear in the IR band
(F160W). The F547M image from Program GO-12174  
has only one pointing
and the smallest sky coverage. The observation in the 
F435W band covers most of the
field-of-view, except for the southwest of the M31 bulge. When we perform 
the SED fitting in \S\ref{s:analysis}, we use all available
filters. All 
pixels over the field-of-view of the F547M image have ten 
data points (except for the `death star' regions of the
\wfc3/IR and the \wfc3/UVIS CCD gap). However, 
the bottom-right (southwest) corner is covered by only six PHAT
filters. 

\subsection{FIR Dust Mass Surface Density Map}\label{ss:far}
We updated the FIR dust mass surface density map of~\citet{gro12} to
be contrasted with our map derived from the extinction. New calibration files 
were applied and the
absorption cross-section was updated to the value given
in~\citet{dra13}. A $\chi^2$ minimization SED fitting with a 
modified blackbody was employed 
on three \herschel\ bands: \pacs\ 100 $\mu$m, 160 
$\mu$m and \spire\ 250 $\mu$m to measure the dust temperature 
and mass, as well as the emissivity index, $\beta$. 
The pixel size of this surface density map is 6\arcsec , while the
FWHM is limited by the \spire\ 250 $\mu$m camera, 18\arcsec . 

We first compare the map of~\citet{gro12} with other three recent ones: two 
from~\citet{dra14} and one from~\citet{via14}.~\citet{dra14} 
fitted a physical dust SED model to the same \herschel\ dataset 
and \spitzer\ observations from 3 to 500 $\mu$m. They provided two 
versions of the dust mass surface density map, with or without
\spitzer/\mips\ 160 $\mu$m and \herschel/\spire\ 500 $\mu$m 
data included in the fit. The
former map has a higher angular resolution (25\arcsec ), while the latter 
one, though with poorer resolution (39\arcsec ), allows for better 
constraint on the dust with temperature below 20 $K$.~\citet{via14}
modelled the far-UV to FIR SED constructed 
with \galex , SDSS, WISE, \spitzer\ and \herschel\
observations. The size of their fitting element is
36\arcsec$\times$36\arcsec . Fig.~\ref{f:dust_self_compare} compares 
the enclosed dust mass as a function of galactocentric radius derived 
from these four maps. The two curves from the maps of~\citet{dra14} are 
very similar because the M31 bulge is dominated by hot 
dust~\citep[$\sim$35 $K$,][]{dra14,gro12}. The dust mass surface
densities of~\citet{dra14} 
and~\citet{gro12} differ by as much as 100\% in the central 50\arcsec\ 
radius, but converge beyond that. Therefore, the total mass within
the central 180\arcsec\ from the map of~\citet{dra14} is only larger
than that of~\citet{gro12} by 18\%. The dust mass derived 
by~\citet{via14} is smaller than those of~\citet{dra14} 
and~\citet{gro12}; the total mass within
the central 180\arcsec\ from the map of~\citet{via14} is only 65\% of 
that of~\citet{gro12}. We suspect that this shortfall is a result of the
shallower observations or different calibration procedures used 
by~\citet{via14}. Based on these comparisons, 
we suggest that the uncertainty in the FIR dust mass 
surface density map is $\sim$50\%.

\section{Analysis and Results}\label{s:analysis}
\subsection{Method}\label{ss:method}
\begin{figure*}[!thb]
  \centerline{
       \epsfig{figure=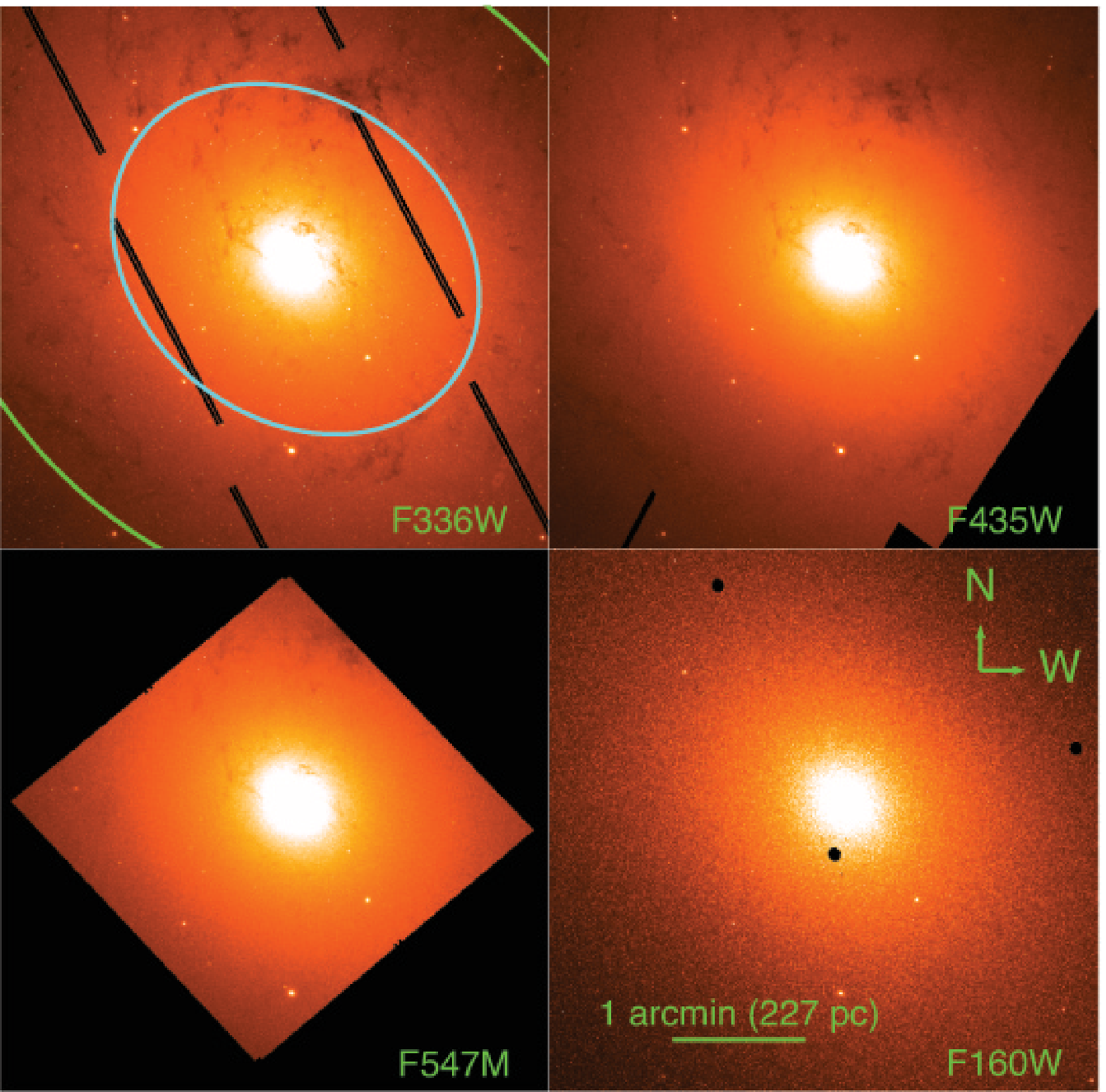,width=1.0\textwidth,angle=0}
       }
 \caption{The mosaic images of F336W, F435W, F547M and F160W. The
dusty clumps in the CNR of M31 can be clearly seen as 
the dark fuzzy filaments in the UV band (F336W), appear
 weak in the optical band (F435W/F547M) and disappear in
the IR band (F160W). The black strips in the
 F336W mosaic are the regions which are just covered by one
 dithered exposure and are simply removed to avoid the
 contamination from cosmic-rays. The black dots 
 in the F160W mosaic are the due to`death star' feature in the \wfc3 /IR
 camera. The green lines in the top left panel represent parts of an
 intensity isophote with
 the major-axis radius of 180\arcsec , while the cyan ellipse
 represents the isophote with the major-axis radius of 90\arcsec . }
 \label{f:mosaic}
 \end{figure*}

\begin{figure*}[!thb]
  \centerline{
       \epsfig{figure=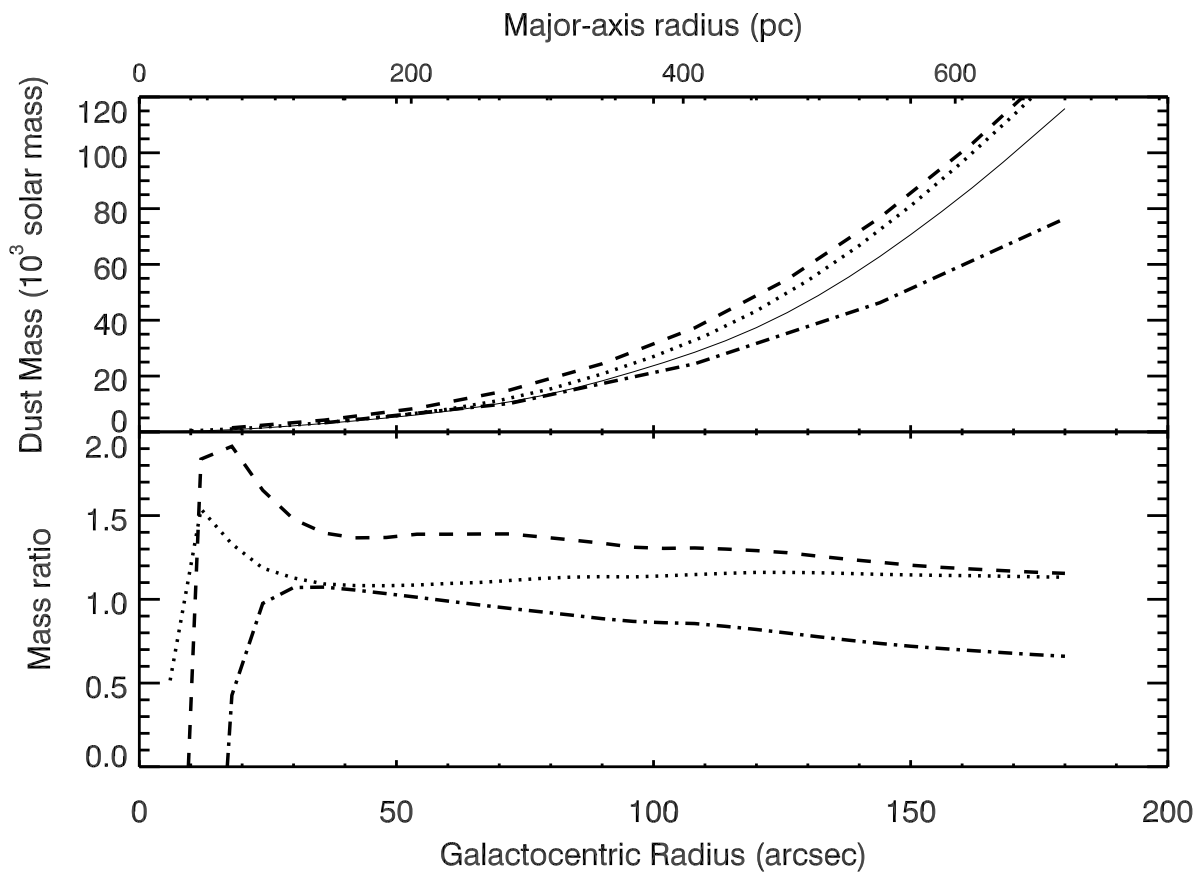,width=1.0\textwidth,angle=0}
       }
 \caption{Top Panel: the enclosed dust mass as a function of the
   galactocentric radius estimated from the four dust mass surface density
   map:~\citet{gro12} (solid), low/high resolution versions
   in~\citet{dra14} (dashed/dotted) and~\citet{via14} (dash
   dot). Bottom Panel: the estimated enclosed dust mass ratios 
of~\citet{dra14} and~\citet{via14} to~\citet{gro12}. 
The rest is the same as those in the
   top panel.}
 \label{f:dust_self_compare}
 \end{figure*}

We utilize a pixel-by-pixel SED fitting method to 
map the extinction distribution in the CNR of
M31 with the \hst\ mosaic images described
above. Considering the relatively 
low column density and small filling factors of dusty 
clumps in the M31 bulge~\citep[][and Fig.~\ref{f:mosaic}]{mel11}, we assume that 
their sizes are  
much smaller than that of 
our covered CNR field and that they are thin and are not 
overlapped with each other along any line of sight. 
Under this assumption, the starlight 
behind a dusty clump is attenuated, 
while its foreground does not.   

\citet{don15} focused on constraining the stellar populations, 
using a $\chi^2$ minimization. The $\chi^2$ for each
pixel is 
defined as the sum of  the deviations between the observed
flux $I_n$ (in the n$^{th}$ filter) and the theoretical flux $S_n$ of a
`Starburst99' stellar population synthesis model~\citep{vaz05} with certain age and metallicity: 
\begin{equation}\label{e:chi}
\chi^2=\sum^{N}_{n=1}\frac{\langle
  I_n-[(1-f)+f\times10^{-0.4\times\frac{A_{n}}{A_{F547M}}A_{F547M}}]S_n\aleph\rangle^2}{\sigma_n^2}
\end{equation}
where 0$\leq$$f$$\leq$1 is the
fraction of the starlight that is subject to the extinction, $A_{F547M}$ is the absolute
extinction in the F547M band, $A_{n}$/$A_{F547M}$ is the
relative extinction, $\sigma_n$ is the 
uncertainty of $I_n$ (see \S\ref{s:observation}) and $\aleph$ is a normalization. 
The sum is over all
of the filters that cover the pixel. We adopt the average extinction
curve, $A_n$/$A_{F547M}$, of five well-defined dense dusty
clumps in the CNR of M31, as obtained by~\citet{don14}, given in
Table~\ref{t:obs}. $A_{F547M}$ is not constrained to be positive, 
so that we can derive the uncertainty of $A_{F547M}$ of our SED
fitting from the $A_{F547M}$  distribution later. 

To break the notorious
extinction-age-metallicity degeneracy, we 
determined $S_n$ independently. As detailed in~\citet{don15}, we 
first performed the SED fitting for several regions of 
low-extinction in the
southeast portion of the M31 bulge, leading to a characterization of  the radial
distribution of the stellar age and metallicity, out to a 
projected major-axis radius of
180\arcsec\ (see Fig.~\ref{f:mosaic}). We used two
  single-epoch populations and fit the observed magnitudes from the
  ten bands with six free parameters: the ages and metallicities of
  the two populations, the mass fraction of the young stellar
  population and the total mass. One starburst represents the 
well-established old stellar population ($\sim$12 Gyr), while 
the other has an age of a few hundred Myr (0.3$-$1 Gyr). 
Both stellar populations are metal-rich. 
The young stellar population contributes $\sim$1\% of the 
total stellar mass in the inner bulge of M31.
This radial SED distribution was then used to calculate 
$S_{n}$ for pixels in the individual isophotes, which nicely follow an elliptical
shape. The pixels outside the outermost isophote, 
corresponding to the outermost radius of the SED distribution
measurement, 
were not used in the subsequent fit.

The fit is to obtain A$_{F547M}$ and $f$, as well as the normalization $\aleph$. 
Compared to previous work~\citep{mel00,mel13}, $f$ is a new 
parameter and is anti-correlated with $A_{F547M}$. 
Because of this anti-correlation, the uncertainties of these two 
parameters can be large for any pixel with the ratio of the 
observed to intrinsic intensities smaller than 1 by less than
two-sigma in the near-UV and optical bands (see~\citealt{don14} for more
details about the intrinsic intensities and the sigma), 
due to either a low $A_{F547M}$ or a small 
fraction of obscured starlight $f$. To overcome this coupling, we 
assume that $f$ changes smoothly across the field 
(e.g., dusty gas forms a coherent structure in the bulge, 
the consistency of which is checked later). Specifically, we 
smooth the $f$ map from the initial $\chi^2$ fit 
with a 11$\times$11 pixel box median filter to reduce the pixel-by-pixel 
fluctuation and to increase the reliability of the $f$ estimation for
the pixels with either low extinction or a small fraction of
obscured starlight. 


After $f$ is derived and fixed, we rerun the $\chi^2$ 
fitting to determine $A_{F547M}$ 
and $\aleph$ for each pixel. The peak of reduced $\chi^2$ is around
0.46 for all pixels and is less than 2 for 81\% of the individual pixels, 
indicating an acceptable fit. 

\subsection{Extinction Map}
\begin{figure*}[!thb]
  \centerline{
      \epsfig{figure=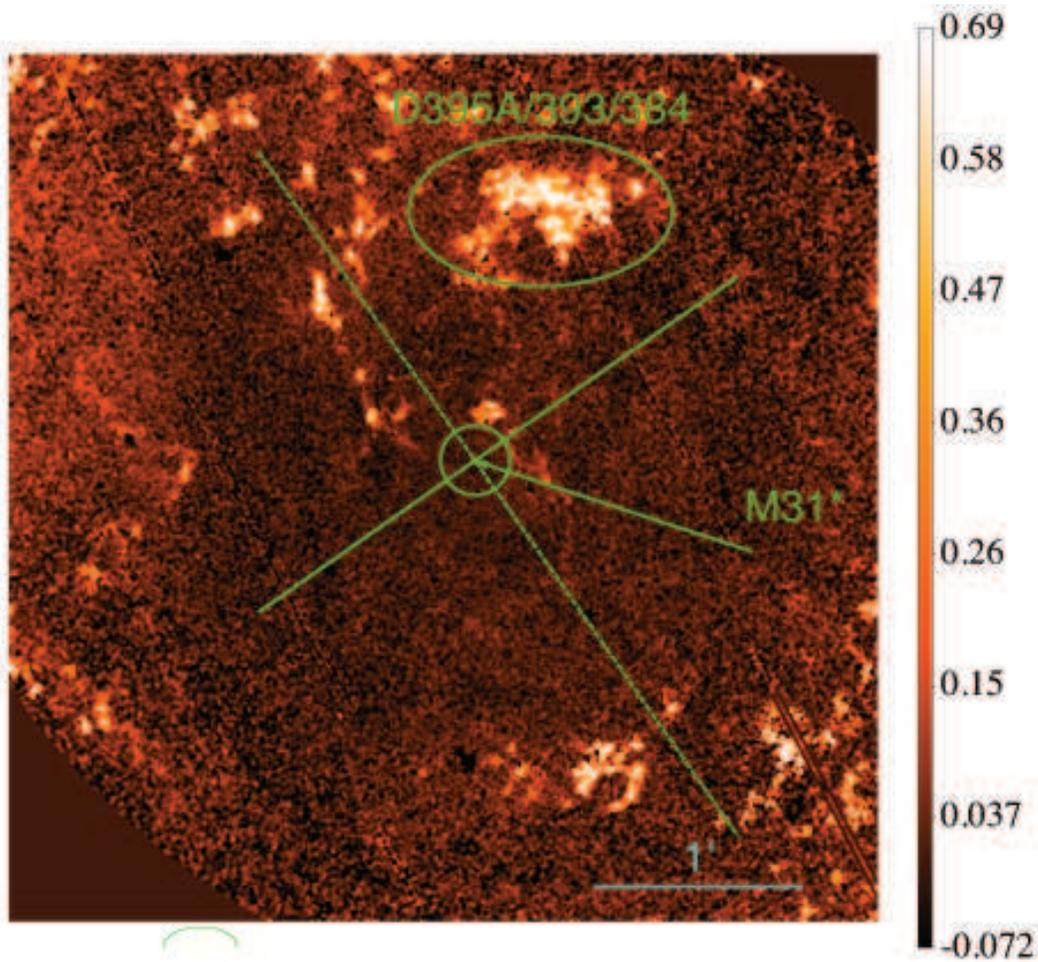,width=1.0\textwidth,angle=0}
}
 \caption{The spatial distribution of the extinction ($A_{547M}$) in the central
   250\arcsec$\times$250\arcsec\ region of M31. The green circle shows the
central 10\arcsec\ around M31* and the arrow points to M31*; the
dashed and solid green lines represent the major and
 minor axes of the M31 bulge; the green ellipse
 outlines D395A/393/384 studied by~\citet{mel00}.}
 \label{f:extin_solo}
 \end{figure*}

Our final extinction map  ($A_{F547M}$) is presented in
Fig.~\ref{f:extin_solo} with color bar, in which 
bright regions represent dusty clumps with high
extinction. Overall, the
extinction map is very smooth and we do not see any artifacts 
introduced by the different numbers of bands used in the
SED fitting across the map.

\begin{figure*}[!thb]
  \centerline{
      \epsfig{figure=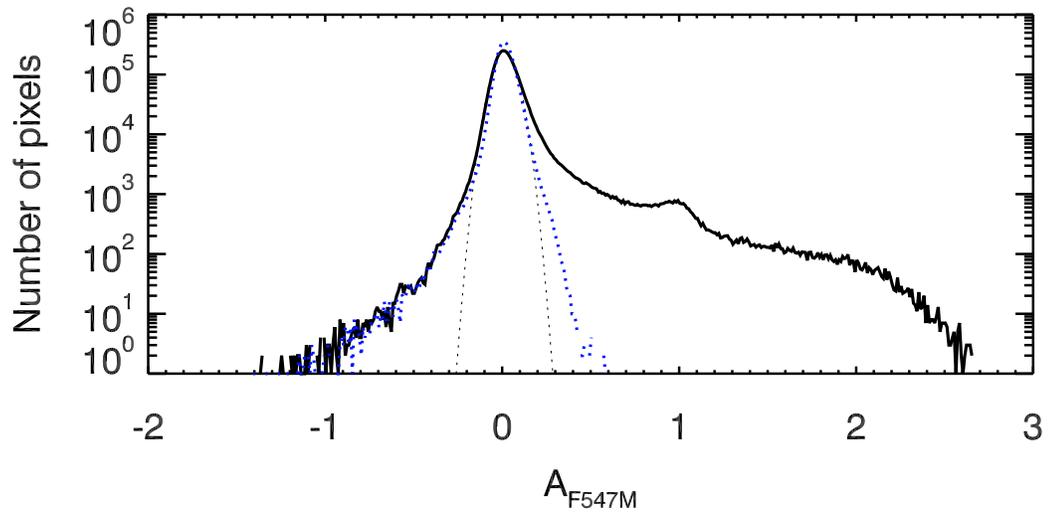,width=1.0\textwidth,angle=0}
}
 \caption{$A_{F547M}$ distribution with a binsize of 0.01
   mag. The black solid and blue dotted curves are for the extinction maps
   with spatially-varying $f$ and fixed $f$=1, respectively. The latter  
   assumes that all the
   dusty clumps are in front of the M31 bulge. The black curve
 near zero can be well characterized with a
   Gaussian function which centers at 0.014 and has a dispersion
 of $\delta A_{F547M}$=0.055 (black dotted line). }
 \label{f:av_dis}
 \end{figure*}

Fig.~\ref{f:av_dis} presents the distribution of $A_{F547M}$ for 
the fitted pixels (black solid line). The peak of the distribution can 
well be approximated with a Gaussian. Its width is chiefly due to
 the photometric uncertainty, as well as variation in the extinction
 by diffuse dust. Its mean extinction value may be characterized by the 
peak position of the Gaussian at 0.014 mag and the uncertainty 
of our extinction map ($\delta A_{F547M}$) 
by the standard deviation, 0.054 mag. The tail on
the negative side of the extinction distribution is primarily due to 
the presence of  globular star clusters and foreground Galactic halo stars, which are bluer
than the predominant low-mass stars in the M31 bulge. 
The tail on the positive side primarily traces dense dusty clumps in the CNR of
M31. For comparison, Fig.~\ref{f:av_dis} also shows the distribution
 of $A_{F547M}$ obtained with fixed $f$=1 (blue
solid line), which assumes that all dust extinction is in the
foreground of the M31 bulge. We can see that 
$A_{F547M}$ in this alternative distribution 
is underestimated by a
factor of 2-3, compared with that allowing for spatially-varying $f$. 

The filling factor of the dusty clumps in the M31
bulge is very low; only 7\% of the pixels in our field-of-view have  
$A_{F547M}$ larger than 3$\delta A_{F547M}$. Especially in the central
10\arcsec , the extinction
is small ($A_{F547M}$$<$$\delta A_{F547M}$ for
94\% pixels), which indicates a paucity of dust 
in the nucleus of M31, consistent with
previous works~\citep{li09,mel11}.  

\subsection{$f$ Map}
\begin{figure*}[!thb]
  \centerline{
      \epsfig{figure=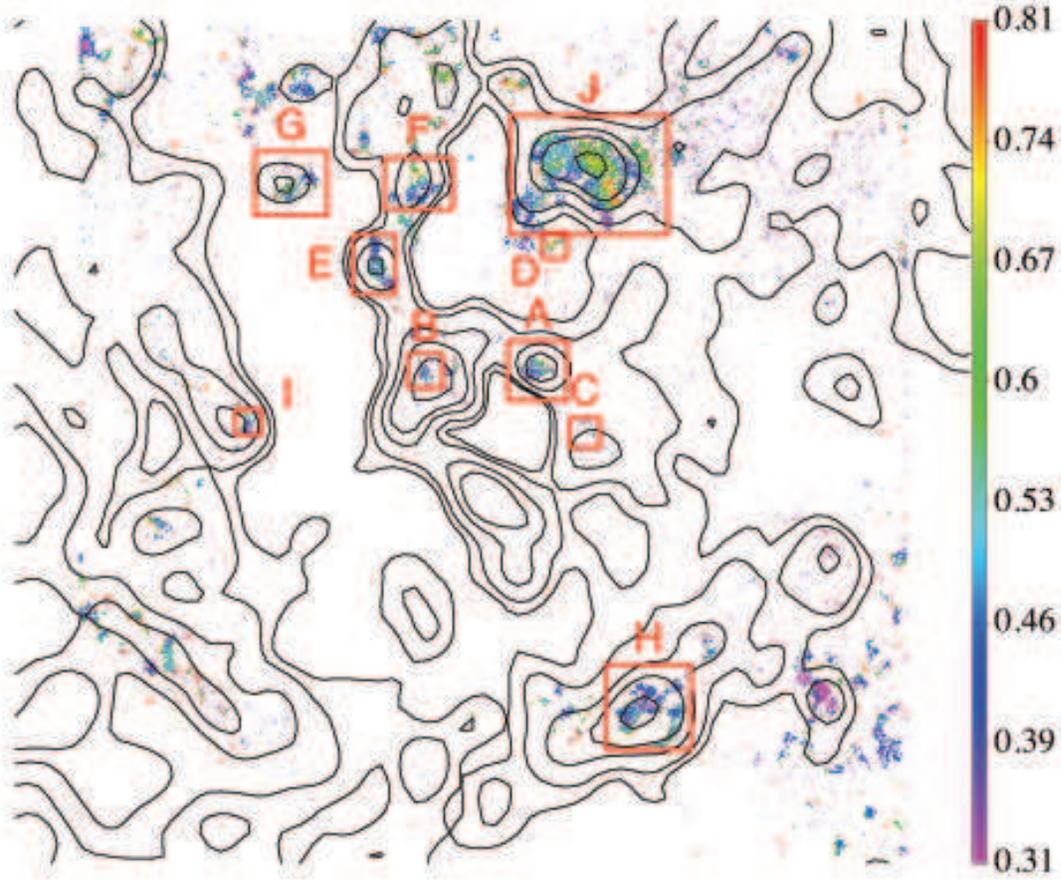,width=1.0\textwidth,angle=0}
}
 \caption{The spatial distribution of the
 fraction of obscured starlight, $f$, for the pixels with
 $A_{F547M}$$>$3$\delta A_{F547M}$, overlaid with the contours 
from \spitzer/\irac\ 8 $\mu$m `dust-only' image. Regions, `A' to 'J', 
are selected to calculate the mean and standard deviation of
$f$, which are listed in Table~\ref{t:f}.}
 \label{f:frac}
 \end{figure*}

Fig.~\ref{f:frac} shows the spatial
distribution of $f$ for the pixels with
$A_{F547M}$$>$3$\delta A_{F547M}$, which have small uncertainty in 
$f$. For most of these pixels, the ratio of the observed to intrinsic
intensities is smaller than 1 by more than two-sigma 
in the near-UV and optical bands 
and did not experience the smoothing step of $f$ 
(\S\ref{ss:method}). Table~\ref{t:f} includes the mean and
standard deviation of $f$ for such pixels in ten selected 
dusty clumps shown in Fig.~\ref{f:frac}. The
dusty clumps `A' to `E' are used in~\citet{don14} to constrain the
extinction curve in the M31 bulge. The fractions of obscured 
starlight of these dusty
clumps from~\citet{don14} and this work are consistent with each other
within twice the measurement uncertainties.  

We follow the steps in Appendix C of~\citet{don14} to translate the 
fraction of obscured starlight
into the line-of-sight offsets of the dusty clumps from M31*. 
This translation assumes a two-dimensional 
Sersic profile for the the M31 bulge 
with a fitted index of 2.2 (Li et al. 2016, in preparation; 
see also~\citealt{kor99,cou11}). Table~\ref{t:f} 
includes the derived line-of-sight offsets and the projected off-center distances 
of the clumps.

\subsection{Dust Mass}\label{ss:dust}
We convert our extinction map to the dust mass surface density
map ($\sum$M$_{de}$ in units of M$_{\odot}$/pixel) by choosing the 
ratio of $\frac{A_{F547M}}{M_{dust}}$. 
This ratio assumes a size distribution of dust
grains, which is characterized by the total to selective extinction
ratio, $R_V$. Dong et al. (2014) shows that $R_V$=2.4-2.5 in the M31
bulge is similar to $R_V\sim$ 2.7 in the Small Magellanic Cloud, but much
smaller than the canonical value of 
$\sim$3.1 in the Milky Way. Therefore, we adopt 
$\frac{A_{F547M}}{M_{dust}}$=3.8  (mag M$_{\odot}^{-1}$ pc$^2$), typical of the Small Magellanic 
Cloud~\citep{dra07,dra14}. 
The total dust mass covered by the pixels with $A_{F547M}>\delta A_{F547M}$ in the
central 180\arcsec\ (i.e., 680 pc) region is
$\sim$1.1$\times10^4$ M$_{\odot}$. 
Table~\ref{t:mass} gives the dust mass of the ten clumps shown in
Fig.~\ref{f:frac}. 

\begin{figure*}[!thb]
  \centerline{
      \epsfig{figure=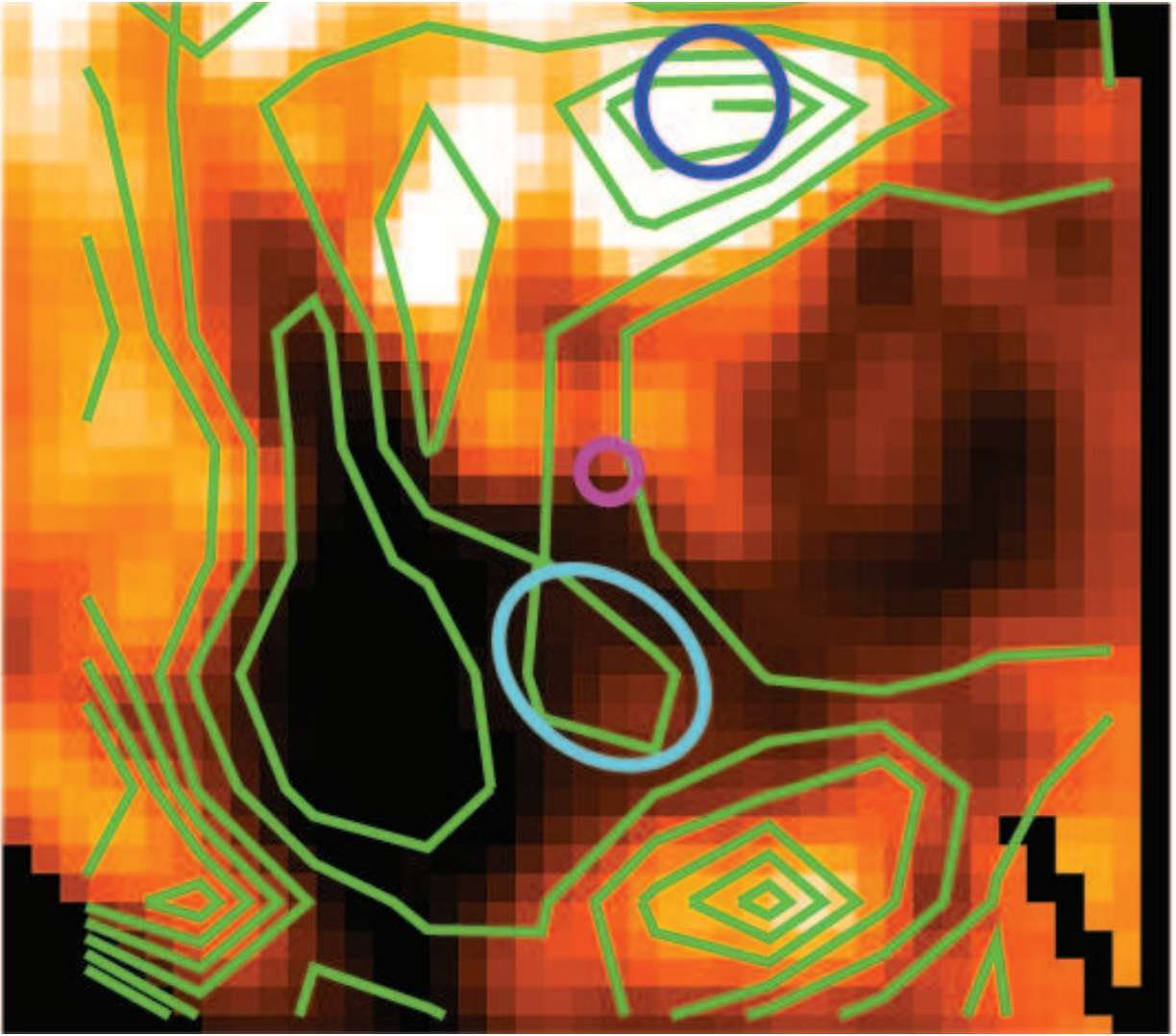,width=1\textwidth,angle=0}
}
 \caption{The dust mass surface density map from our
 extinction map, with a pixel size of 6\arcsec\ (color image), 
compared with the (green) contours 
 from the dust mass map of~\citet{gro12}. The magenta circle
 marks the center of M31 and the blue circle 
indicates D395A/D393/D384. The two dust maps appear similar in
morphology, showing dust filaments and low-extinction patches to both 
east and west of M31*. The cyan ellipse indicates
one dusty clump, 30\arcsec\ (i.e., $\sim$120 pc) southeast of 
M31*, which has been detected in the CO observations, but should be 
on the far side of the M31 bulge~\citep{mel13}.}
\label{f:dust_com}
\end{figure*}

\begin{figure*}[!thb]
  \centerline{
      \epsfig{figure=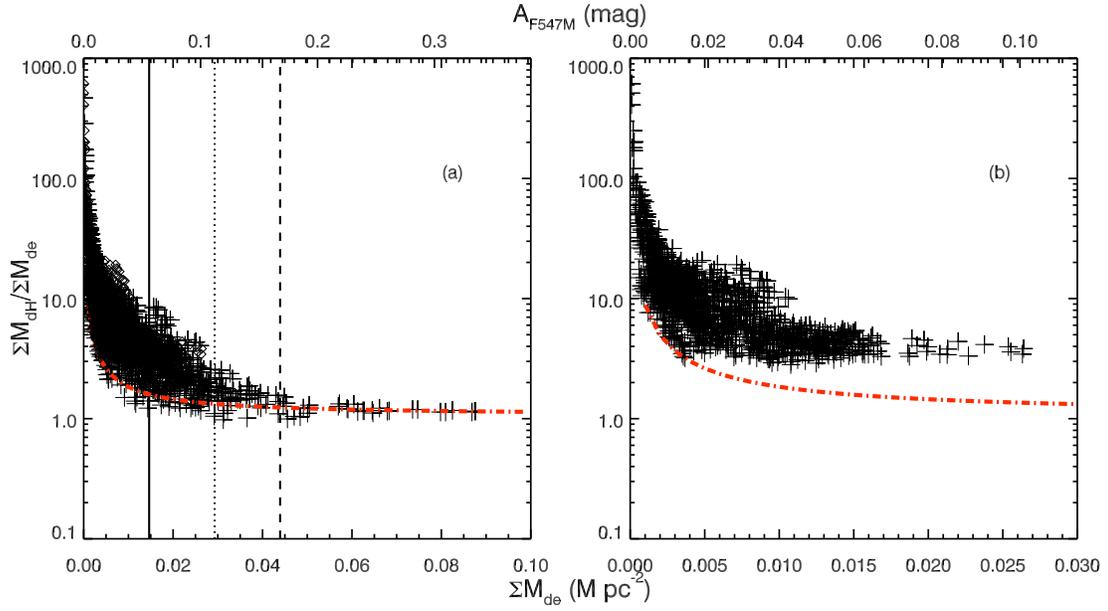,width=1.0\textwidth,angle=0}
}
 \caption{The pixel-by-pixel dust mass surface density 
ratio of the \herschel\ data ($\sum$M$_{dH}$) to our
extinction-inferred 
   map ($\sum$M$_{de}$) with (a) spatially-varying $f$ and (b) fixed $f$=1
   as a function of $\sum$M$_{de}$. 
The abscissa axis ranges of these two plots are different. In
panel (a), the solid, dotted and dashed lines represent one, two and three
 times the uncertainty of our map. 
The red dot dashed lines in both panels represent the same 
linear fit between 
$\sum$M$_{dH}$ and $\sum$M$_{de}$ for pixels 
with M$_{de}$$>$3$\delta$M$_{de}$ in the 
map with spatially-varying $f$ (see \S~\ref{ss:herschel}).}
\label{f:dust_com_pix}
\end{figure*}

We compare our dust mass surface density map  
with that ($\sum$M$_{dH}$) derived from the \herschel\ FIR 
observations described in \S\ref{ss:far}. Because the resolution of the latter  
is much poorer, we convolve 
our map with the PSF of \herschel\ \spire\ 250 $\mu$m and 
then rebin it to 6\arcsec\ pixel$^{-1}$. 
Fig.~\ref{f:dust_com} shows that the two maps 
are broadly similar in morphology. However, 
there are also marked differences between the two maps. For
example, the dusty clump southeast of the M31 bulge, 
as outlined by the cyan ellipse in Fig.~\ref{f:dust_com}, is visible only in far-IR
 (see \S\ref{s:discussion} for more
discussion about this clump). Fig.~\ref{f:dust_com_pix}(a) gives 
a pixel by pixel comparison of
$\sum$M$_{de}$ and $\sum$M$_{dH}$. At pixels of 
low extinction, especially those below its 
uncertainty ($\delta A_{F547M}$), $\sum$M$_{dH}$ is always greater than 
$\sum$M$_{de}$; a good example of this trend is the
dusty clump southeast of the M31 bulge mentioned above. The ratio between 
$\sum$M$_{dH}$ and $\sum$M$_{de}$ decreases with the
increase of $\sum$M$_{de}$. At pixels with 
$\sum$M$_{de}$ above twice the uncertainty,
the ratio is close to 1. Fig.~\ref{f:dust_com_pix}(b) provides 
a similar comparison, but for a fixed $f$=1. 

We also measure the dust-to-gas mass ratio 
in the M31 bulge. The largest dusty 
clump in our field-of-view is D395A/393/384 
(green ellipse in Fig.~\ref{f:extin_solo}), 1.3\arcmin\ 
($\sim$300 pc) northeast of M31*.~\citet{mel11} give
$I_{CO(1-0)}$=0.72$\pm$0.03 K km s$^{-1}$ (with a 24\arcsec\ beam 
size) for this clump, i.e., (9.8$\pm$0.4)$\times$10$^3$ $M_{\odot}$ 
molecular gas, assuming the conversion factor between the CO emission 
and $H_2$ mass given in~\citet{ler11}. In the same region, the
\ion{H}{1} mass is 1.4$\times10^4$ $M_{\odot}$, according to the 
map of~\citet{bra09}. Our surface density map of dust mass gives
706$\pm$219 $M_{\odot}$, which is consistent with the value, 603
$M_{\odot}$, from the FIR map given in \S\ref{ss:far}. Therefore, we
derive a dust-to-gas mass ratio of 0.03$\pm$0.009,  
consistent with the value of 0.026 given in~\citet{dra14} for the entire
center of M31. 

\section{Discussion}\label{s:discussion}
Here, we explore the structure of the dusty materials in the central
region of M31 using our newly estimated locations of the
dusty clumps in \S\ref{ss:f}  and discuss the reliability of the method 
in mapping the dust in \S\ref{ss:herschel}. 

\begin{figure*}[!thb]
  \centerline{
      \epsfig{figure=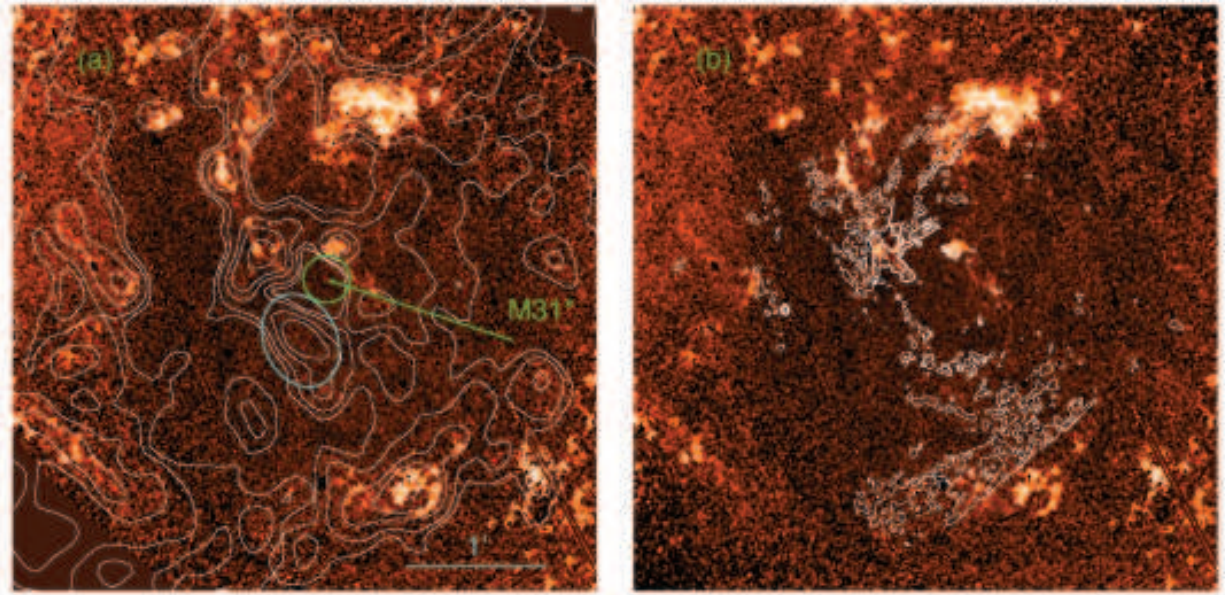,width=1.0\textwidth,angle=0}
}
 \caption{The spatial distribution of the extinction ($A_{547M}$) in the central
   250\arcsec$\times$250\arcsec\ region of M31, overlaid with the contours
   from (a) \spitzer/\irac\ 8.0 $\mu$m `dust-only' image and (b) 
   H$\alpha$+[\ion{N}{2}] emission from HST imaging (Li et al. 2016, in
   preparation). In (a), the green circle shows the
central 10\arcsec\ around M31*; the cyan ellipse to the
 southeast of M31* marks a dusty clump detected in the
 \spitzer/\irac\ 8 $\mu$m observations, but without any apparent extinction
 of the starlight from the M31 bulge.}
 \label{f:extin}
 \end{figure*}

We first compare our results with other observations. 
The overall structure of our extinction map is qualitatively similar to that 
of~\citet[][]{mel00} (their Figure 1; though at 1\arcsec\ resolution), which is 
derived from a ground-based $B$-band image. The high extinction
regions show filament structures around M31* and trace the 
\spitzer/\irac\ 8 $\mu$m emission very well (Fig.~\ref{f:extin}a),
indicating that the hot dust is indeed associated with the
dusty clumps. As noted by~\citet{mel13}, however, one 
clump southeast of M31* (cyan ellipse in Fig.~\ref{f:extin}a) 
has strong 8 $\mu$m emission, but does not show up in the extinction 
map, indicating that it is located at the far side of the M31
bulge. Fig.~\ref{f:extin}b shows that the H$\alpha$+[\ion{N}{2}]
emission, 
which will be studied in 
details in an upcoming paper by Li et al. (2016, in preparation). 



\subsection{Nuclear Spiral Structure}\label{ss:f}

Which dusty clumps are associated with the nuclear spiral of M31? 
Most of the dusty clumps shown in  Fig.~\ref{f:frac} have the 
radial (line-of-sight) offsets from M31*
smaller than 50 pc, which indicates that they are indeed within the
bulge. This provides a natural explanation of their  
associations with the 
ionized gas (H$\alpha$+[\ion{N}{2}]), which is thought to be 
heated by hot stars concentrated toward the nucleus 
(e.g., post-AGBs;~\citealt{bin94}) and by type-Ia
supernova shocks in the bulge~\citep{jac85}. 
In contrast, Clump D, and the 
clump enclosed by the cyan ellipse in Fig.~\ref{f:extin}a do not
show significant H$\alpha$+[\ion{N}{2}] emission. Therefore, these two
dusty clumps likely have large off-nucleus radial distances.

 Because of their close spatial distribution, these ten dusty clumps 
may belong to a coherent 
disk-like structure, only slightly inclined with respect to our line of
sight. The clumps to the northeast, such as `B', `E' and `F', are 
slightly beyond the nucleus 
(but still $<$25 pc). These dusty clumps extend northwest to 
`D395A/393/384' (`J' in Fig.~\ref{f:frac}) , which becomes in
front of the bulge. On the other hand, the dusty clumps in the south, 
`H' and `I' are behind M31*. 

\begin{figure*}[!thb]
  \centerline{
      \epsfig{figure=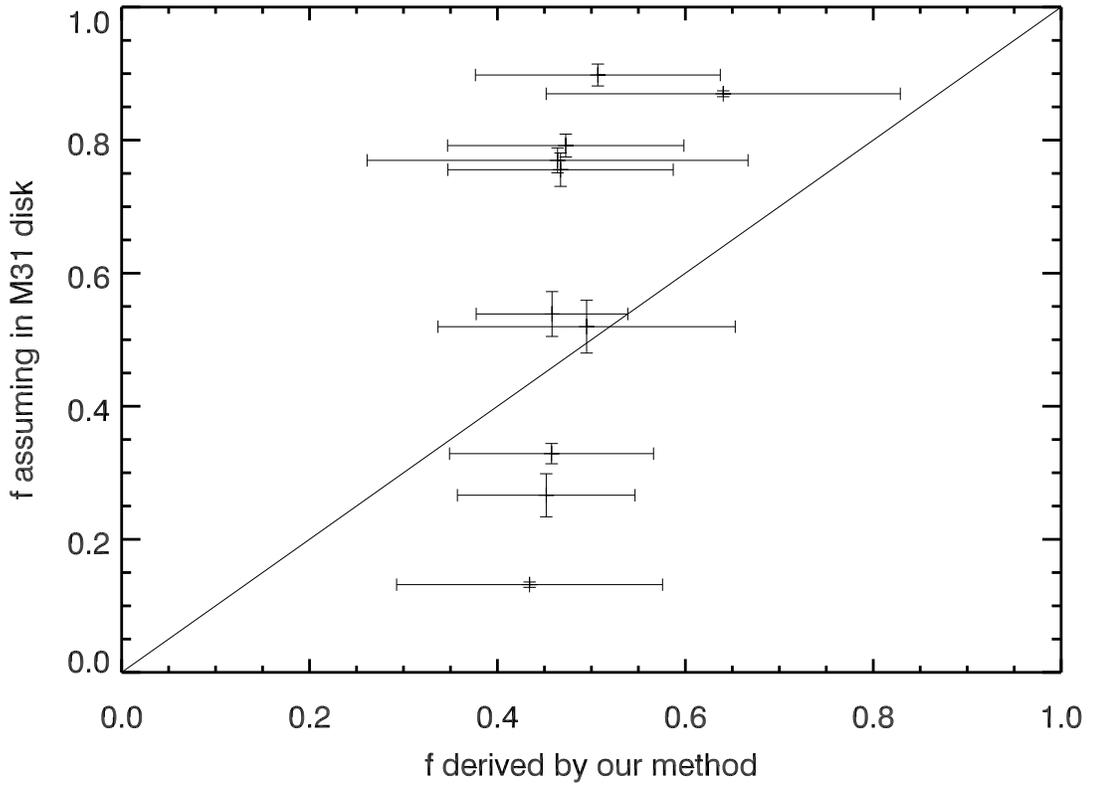,width=1.0\textwidth,angle=0}
}
 \caption{Comparison of the obscured fraction $f$ for the dusty clumps in
   Fig.~\ref{f:frac} between the cases of using our method and the
   case of assuming 
that they are located in the same plane as the M31 disk.}
 \label{f:My_Lzy}
 \end{figure*}

This structure does not align with the galactic disk of M31, which has
an inclination angle of 77 deg. 
Figure~\ref{f:My_Lzy} compares the $f$ values for the individual 
dusty clumps in Table~\ref{t:f} with what would be expected if 
they were in the galactic disk plane. This expectation is based on 
the starlight model of M31, in which the stellar 
bulge is a triaxial ellipsoid and that the 
M31 disk is a thin plane. The model $f$ values are then derived 
from their projected off-nucleus distances, following the procedure 
detailed in Appendix C of~\citet{don14}. Figure~\ref{f:My_Lzy}, 
 shows that the range of the $f$ values (all around 0.5) 
is much smaller than expected from the galactic disk model. 
Therefore, the nuclear dusty structure is tilted off from the galactic
disk and is consistent with a nearly face on configuration.


The misalignment of the nuclear dusty structure and the galactic disk 
indicates violent activities which happened in the M31 bulge in the
recent past. 
The dusty clumps 
in the bulge of spiral galaxies can originate from the disk and be 
transported into the bulge through the torque of a bar~\citep{kor04}.  
Therefore, the dusty clumps are supposed to inherit the dynamical 
information of the galactic disk. External forces are 
needed to disturb the dusty clumps from their original orbit. 
The head-on collision between M31 and M32 suggested by
\citet{blo06} and ~\citet{mel11} could 
play such a role in shaping the molecular clouds in the CNR of
M31. 

\subsection{The Reliability of Tracing Dust with
  Extinction}\label{ss:herschel}

\citet{kre13} discuss the reliability and feasibility of mapping dust with
extinction estimated from the Balmer decrement and 
stellar continuum shape for eight
nearby galaxies. From the intensity ratio of
H$\alpha$/H$\beta$ and the SED-fitting of the observed stellar 
continuum in the wavelength range of 3700$-$7000 \AA , they 
produce two apparent dust reddening
maps with resolutions of 20-100 pc for these
galaxies. However, the 
extinction derived from 
the stellar continuum shows no linear correlation with the dust mass 
from the \herschel\ FIR data, which is considered as a standard. 
This nonlinearity is attributed to the mixed 
distribution between the dust and the light from stars 
with a board range of ages. In contrast, a linear correlation is 
found between the dust mass and the extinction derived from 
the line ratio, which can be explained by a close association of 
the dense dusty gas with recent/ongoing star-forming regions. 
 
\begin{figure*}[!thb]
  \centerline{
      \epsfig{figure=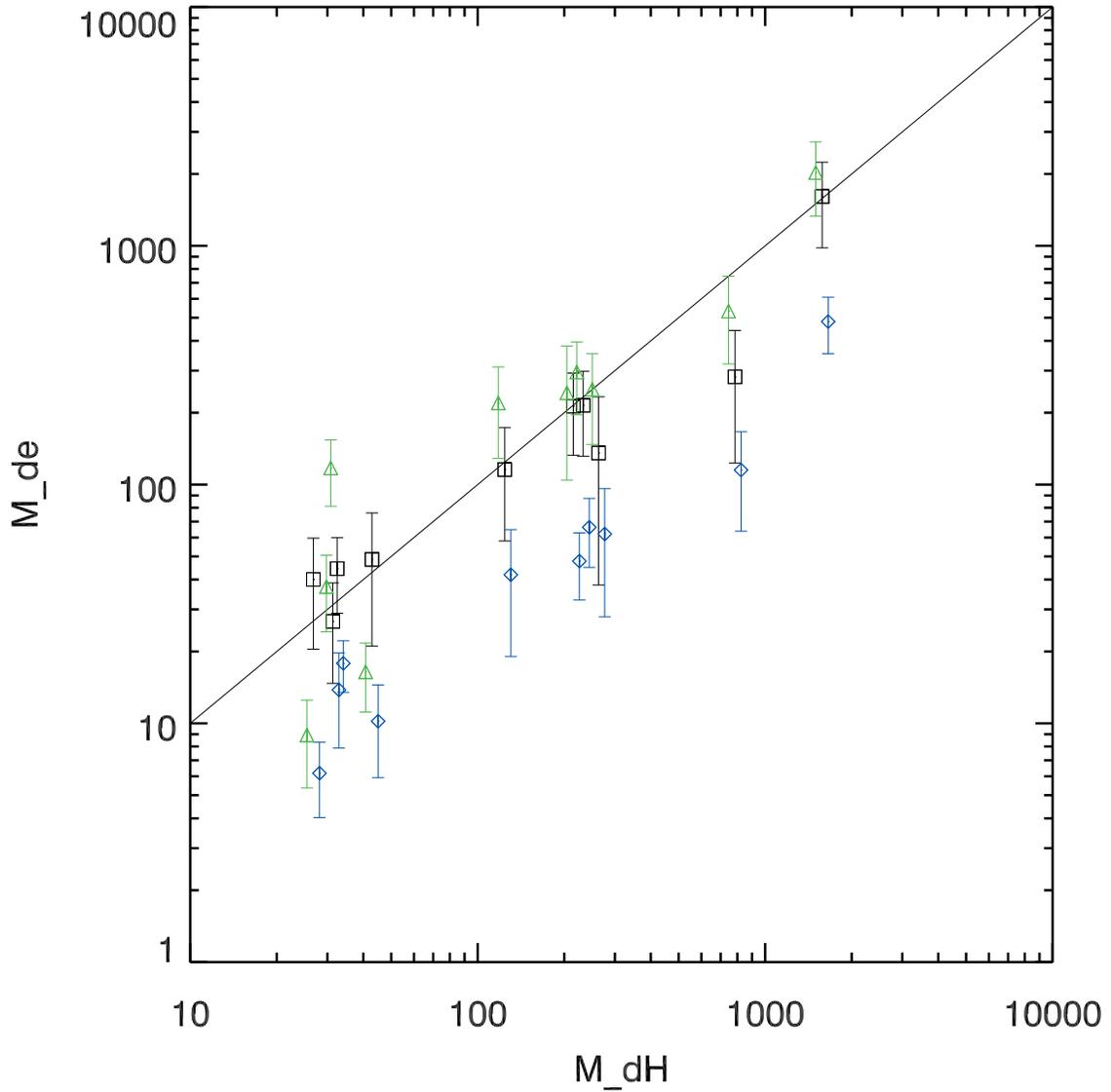,width=1.0\textwidth,angle=0}
}
 \caption{Comparison of the masses of the dusty clumps (in units of M$_{\odot}$) in
   Fig.~\ref{f:frac} from the extinction map ($M_{de}$) with spatially-varying
   $f$ (black squares) or fixed $f$=1 (blue diamonds) with those from
   the \herschel\ data ($M_{dH}$). The green triangles represent the
   masses of clumps, assuming that they were in the Virgo cluster,
   derived by our method with spatially-varying $f$. The blue diamonds
   and green triangles have been slightly shifted along the abscissa
   axis for better demonstration.}
\label{f:dusty_clump_compare}
\end{figure*}

Unlike~\citet{kre13}, we find a linear relationship between the dust
mass derived from the extinction and dust emission in the high extinction
regions in the CNR of M31.  At $A_{F547M}$$>$3$\delta A_{F547M}$,  
$\sum$M$_{dH}$=(1.06$\pm$0.07)$\times\sum$M$_{de}$+(0.008$\pm$0.004). 
In Fig.~\ref{f:dusty_clump_compare}, we show good agreement between 
the dust mass of the dense dusty clumps derived from our method with 
spatially-varying $f$ and for the \herschel\ dat. 
The median and the standard variation of the ratios
between the two masses are 0.99$\pm$0.34. Instead, for fixed $f$=1,
the median and the standard deviation of the ratios would be only
0.28$\pm$0.11. 

This good consistency is due to several advantages in our
work: 1) Our angular
resolution reaches $\sim$ 2 pc (0.5\arcsec ),  better than 
that of~\citet{kre13} by more than an order of magnitude. In the CNR of
M31, even the cores of small dusty clumps seem to be reasonably well
resolved; 
2) The relative distribution between starlight and extinction in the M31
bulge is much 
simpler than in the irregular and spiral galaxies in~\citet{kre13};  
3) Most importantly, in our
SED-fitting we have introduced the parameter $f$, 
which allows us to quantitatively separate the unobscured 
foreground starlight from the background.

\begin{figure*}[!thb]
  \centerline{
      \epsfig{figure=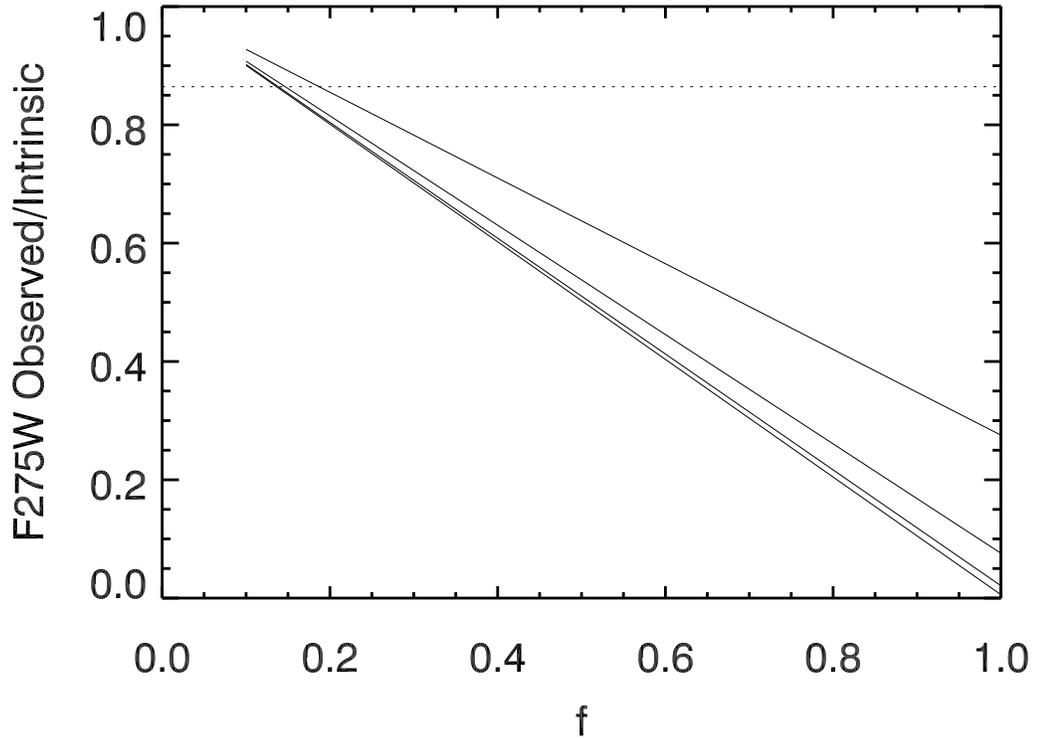,width=1.0\textwidth,angle=0}
}
 \caption{Demonstration of the ratio of the observed to intrinsic
   F275W intensities as a function of $f$ for a molecular cloud with
   $A_{F547M}$=0.5, 1, 1.5 and 2 mag (from top to bottom). The
   horizontal 
 dashed line marks the photometric uncertainty at F275W band for one 
pixel of our mosaic (13.5\%, which consists of
 13.1\% statistical uncertainty and 3.5\% systematic uncertainty, see
 Table 1 of~\citealt{don15}) from 1.}
\label{f:f275w_demon}
\end{figure*}

However, when moving into the low-extinction regions, especially 
for $A_{F547M}<$$\delta A_{F547M}$, we
seriously underestimate the amount of dust (see
Fig.~\ref{f:dust_com_pix}). There are several reasons why the extinction can not
reproduce the dust mass: 1) The dusty clumps, such as the one in 
the southeast of M31* (the cyan ellipse in Fig.~\ref{f:dust_com}), 
are far behind the M31 bulge and hence 
the fractions of obscured starlight are so low that the dusty clumps 
have little signal even in the UV images. This effect is demonstrated 
in Fig.\ref{f:f275w_demon}, which shows the ratio of the observed 
to intrinsic F275W intensities as a function of $f$ for a molecular cloud with
 different $A_{F547M}$. We can see that for $f$$<$0.15, even a cloud
 with $A_{F547M}$=2 can lead to ~14\% extinction, which is comparable
to the photometric uncertainty in this band;
and 2) In the low-extinction regions, the
diffuse dust may pervade the M31 bulge and may be well mixed with 
stars. As a result, they also leave very few signals in the
  observed UV images. From the \herschel\ image, the median dust 
surface density in the M31 bulge is 0.036 M$_{\odot}$/pc$^2$, which corresponds 
to $A_{F547M}$$\sim$0.1. For uniformly distributed diffuse dust, we only 
expect that it reduces the intensity at the F275W band by 
12\%, also smaller than the 
photometric uncertainty. 
Accurately determining the diffuse dust 
needs detailed radiation transfer modeling. 

In the low-extinciton region, the intensity ratio of
H$\alpha$/H$\beta$ also underestimates the dust mass. For example, in
the eight galaxies studied in~\citet{kre13}, the ratio between
the apparent extinction from the stellar continuum and the
H$\alpha$/H$\beta$ is less than one in high extinction regions and
becomes greater than one in low extinction regions. In this latter
case, most of the hydrogen
recombination line emission may arise from the scattered light, which
suffers from less extinction than the stellar continnum. 

We conduct simulations to test how our method may be applied to 
relatively distant galaxies. We artificially 
put M31 at a distance comparable to those of 
the Virgo cluster galaxies and re-measure 
the masses of the dusty clumps in Fig.~\ref{f:frac}. We assume the
distance of the Virgo cluster to be 16.5 Mpc~\citep{ton01},
correspondingly 
downgrade our \hst\ images by a factor of 21, and redo our
measurement as in \S\ref{s:analysis}. The measured dust masses are given in
Table~\ref{t:mass} and are compared with those from the \herschel\ 
observations in Fig.~\ref{f:dusty_clump_compare}. The 
ratio of the dusty masses from the extinction and from the 
far-IR emission is 1.19$\pm$0.9. Although the uncertainty is 
larger than that of 
M31, due to the reduced physical resolution, the median value 
suggests that our method can still give an reasonably accurate estimate of 
the dust mass for molecular clouds in distant galaxies. 

Therefore, extinction measurements could be used to map dust in
remote galaxies, when two conditions are satisfied: 1) The sizes of
molecular clouds are discrete and compact, compared to the 
extent of a galactic spheroid, such as dust lanes in giant elliptical
galaxies, where a screen assumption can reasonably be made; 
2) The fraction of the 
obscured starlight, $f$, should be large
enough that the extinction can be accurately measured. 
In such case, we can then simultaneously determine both 
the fraction of obscured
starlight and the amount of dust.  
 
\section{Summary}\label{s:summarize}
We have mapped the dusty clumps in the circumnuclear region of M31
via their extinction of background starlight in multiwavelength
bands. Thanks to the high angular resolution and sensitivity
of the \wfc3\ and
\acs\ data, we have achieved an unparalleled $\sim$$0.5$\arcsec\ 
(2 pc) resolution in the final extinction map. Because our dataset spans from near-ultraviolet to
near-infrared, the map has a large 
dynamic range and an uncertainty of $\delta A_{F547M}$=0.055
mag. 

The main innovation of our work is the measurement of the 
fraction of obscured starlight, $f$, across the field. 
With the fraction included in our SED fit  for dense dusty clumps, 
the extinction typically increases significantly by a factor of 2-3, 
compared to those by simply
assuming a foreground screen ($f$=1). We have found that most of the
clumps have $f$$\sim$0.5 (line-of-sight offset from M31*, $<$ 50 pc), indicating 
that they are indeed in the central region of M31 and represent a 
coherent structure. Compared to the M31 disk, this structure has a small
inclination and seems to be consistent with the inner ring
claimed by~\citet{mel11}. This geometry agrees with the
prediction from the H$\alpha$-emitting gas observations; the shocked 
heated gas are associated with the inner side of the structure around
M31*, consistent with the scenario 
that the ISM in the center of M31 was strongly disturbed in the recent
past by, for example, a collision between M31 and M32. 

We have converted our extinction map into a surface density
map of dust mass. The total dust mass in D395A/393/384, the 
largest dusty clump in
the central 2\arcmin\ of the M31 bulge, is $\sim$700
$M_{\odot}$. We have also inferred the dust-to-gas mass ratio,
0.03$\pm$0.009, which is consistent with the value given
in~\citet{dra14}. 

We have compared our surface density map with that from
 recent \herschel\ far-infrared observations. In the low
 extinction regions, where our map seriously 
  underestimates the dust mass. In the regions with
  $A_{F547M}~>~3\delta A_{F547M}$, however, the dust extinction 
is consistent with the
  amount of dust, as seen in far-infrared emission. Our simulations
  have further shown that a similar method may be used to 
detect high extinction clumps in early-type 
galaxies in the Virgo cluster. 

\section*{Acknowledgments}
We thank the anonymous referee for a thorough, detailed, and
constructurive commentary on our manuscript. 
This work uses observations made 
with the NASA/ESA Hubble Space Telescope and 
the data archive at the Space Telescope Science Institute, which is 
operated by the Association of Universities for Research in Astronomy, 
Inc. under NASA contract NAS 5-26555. The work is partially supported by 
NASA via the grant GO-12055 and GO-12174 
 provided by the Space Telescope Science Institute. 
 The work has also received funding from the European 
Research Council under the European Union's Seventh 
Framework Programme (FP7/2007-2013) / ERC grant 
agreement n° [614922]. H. D. 
acknowledges the support and hospitality of the Key Laboratory 
of Modern Astronomy and Astrophysics at 
Nanjing University during his visit, and would like to thank Bruce
Draine, Robert Braun and Sebastien Viaene for providing their dust 
mass surface density maps and \ion{H}{1} map. Z. L. acknowledges
support from the Recruitment Program of Global Experts and the
National Science Foundation of China through grant 131147. 


\begin{deluxetable}{cccccccc}
\rotate
  \tabletypesize{\small}
 \tablecolumns{8}
  \tablecaption{\hst\ Multi-Wavelength Observations}
  \tablewidth{0pt}
  \tablehead{
  \colhead{}&
  \colhead{}&
  \colhead{Pivot $\lambda$} &
  \colhead{PHOTFLAM} &
   \colhead{Systematic error} & 
   \colhead{Median of } & 
   \colhead{$A_n/A_{F547M}$} &  
   \colhead{$A_n/A_{F547M}$}\\
  \colhead{Filter} & 
  \colhead{Detector} &
  \colhead{($\AA$)} &
  \colhead{{\rm ergs $cm^{-2}$ $s^{-1}$ $\AA^{-1}$}} &
  \colhead{of PHOTFLAM$^{a}$} & 
  \colhead{$\sigma$/I (\%)$^{b}$} & 
  \colhead{Milky Way$^c$} & 
  \colhead{M31 bulge$^d$} \\
  }
  \startdata
F275W & \wfc3 /UVIS & 2704 & 3.3010e-18 & 3.5\% &
13.1&1.92 & 2.80\\
F336W & \wfc3 /UVIS & 3355  & 1.3129e-18 & 2\% & 3.7 &
1.62 & 1.92\\
F475W & \acs /WFC & 4747 &1.8210e-19 & 2\% & 4.1 &
1.16 & 1.21\\
F814W & \acs /WFC & 8057 & 7.0332e-20 & 2\% & 7.8 &
0.58 & 0.61\\
F110W & \wfc3 /IR & 11534 & 1.5274e-20 & 2\% & 12.7 &
0.33 & 0.46\\
F160W & \wfc3 /IR &  15369 & 1.9276e-20 & 2\% & 14.7
& 0.20 & 0.43\\
F390M & \wfc3 /UVIS & 3897 & 2.5171e-18 & 2\% & 3.1
& 1.48 & 1.58\\
F547M & \wfc3 /UVIS &  5447 & 4.6321e-19 & 2\% &
4.1 & 1.0 & 1.0\\
F665N & \wfc3 /UVIS &  6656 & 1.9943e-18 & 2\% &
5.0 & 0.80 & 0.73\\
F435W & \acs /WFC & 4318.9 & 3.1840e-19 & 2\% &
3.6 & 1.30 & 1.37\\
\enddata
\label{t:obs}
\tablecomments{ a) The `PHOTFLAM' information is obtained from:
  http://www.stsci.edu/hst/wfc3/phot\_zp\_lbn (WFC3) and
  http://www.stsci.edu/hst/acs/analysis/zeropoints/\#tablestart (ACS).
b) `I' and $\sigma$ are the intensity and uncertainty at individual 
  pixels (in unit of electron s$^{-1}$). c) The relative extinction
  $A_n/A_{F547M}$ for the MW-type dust. d) The average
  relative extinction $A_n/A_{F547M}$ of the five dusty clumps in the
  M31 
bulge, derived
  in~\citet{don14}. }
\end{deluxetable}

\begin{deluxetable}{cccc}
  \tabletypesize{\small}
 \tablecolumns{4}
  \tablecaption{$f$ values and off-M31* Distances of the Dusty Clumps}
  \tablewidth{0pt}
  \tablehead{
  \colhead{ID}&
  \colhead{$f$}&
  \colhead{Projected Distance} &
   \colhead{Radial Distance (pc)$^a$} \\
  }
 \startdata
A&0.47$\pm$0.13&13\arcsec (52 pc)&7.4$\pm$26.3\\
B&0.46$\pm$0.11&32\arcsec (123 pc)&16.3$\pm$37.9\\
C&0.46$\pm$0.20&19\arcsec (71 pc)&10.7$\pm$55.3\\
D&0.64$\pm$0.19&51\arcsec (196 pc)&-62.4$\pm$101.6\\
E&0.46$\pm$0.08&64\arcsec (242 pc)&23.0$\pm$41.1\\
F&0.47$\pm$0.12&76\arcsec (289 pc)&20.1$\pm$67.7\\
G&0.49$\pm$0.16&99\arcsec (374 pc)&5.2$\pm$102.1\\
H&0.45$\pm$0.09&95\arcsec (362 pc)&31.7$\pm$59.4\\
I&0.43$\pm$0.14&83\arcsec (314 pc)&40.0$\pm$85.2\\
 J&0.51$\pm$0.13&74\arcsec (283 pc)&-1.7$\pm$72.7\\
 \enddata
\tablecomments{ a) the positive (negative) values mean that the clumps
are behind (in front of) the plane of M31*. }
\label{t:f}
\end{deluxetable}

\begin{deluxetable}{ccccc}
  \tabletypesize{\small}
 \tablecolumns{5}
  \tablecaption{The Mass of Dusty Clumps}
  \tablewidth{0pt}
  \tablehead{
  \colhead{ID}&
  \colhead{M$^a$} &
  \colhead{M$_{f=1}^b$} &
  \colhead{M$_{Herschel}^c$} & 
  \colhead{M$_{Virgo}^d$} \\
  }
  \startdata
A & 116$\pm$58 & 42$\pm$23 & 124 & 220$\pm$91 \\
B & 40$\pm$20  & 6$\pm$2 & 27 & 9$\pm$3 \\
C & 27$\pm$12  & 14$\pm$6 & 31 & 37$\pm$13 \\
D &  44$\pm$16  & 18$\pm$4 & 32 & 117$\pm$36 \\
E &  213$\pm$80  & 48$\pm$15 & 215 & 242$\pm$138 \\
F &  215$\pm$83  & 66$\pm$21 & 233 & 296$\pm$100 \\
G & 136$\pm$98  & 62$\pm$34 & 263 & 251$\pm$103 \\
H & 283$\pm$159  & 115$\pm$51 & 785 & 533$\pm$213 \\
I & 49$\pm$27  & 10$\pm$4 & 43 & 16$\pm$5 \\
 J & 1609$\pm$629  & 482$\pm$128 & 1578 & 2028$\pm$695 \\
 \enddata
\label{t:mass}
\tablecomments{ the dust mass in units of solar mass 
a) derived from the attenuation map with 
spatially varied $f$; b) derived from 
the attenuation map with 
fixed $f$=1; c) measured from the \herschel\ data; d) measured from 
the attenuation map with spatially varied $f$, but assuming that the
clumps were in the
Virgo Cluster. }
\end{deluxetable}



\end{document}